\documentclass{PoS}

\usepackage{bm}
\usepackage{braket}
\usepackage{nicefrac}
\usepackage{pgf}
\usepackage{subcaption}
\usepackage{siunitx}

\usepackage{amsmath}
\usepackage{amssymb}
\usepackage{amsfonts}
\usepackage{amsthm}
\usepackage{bm}
\usepackage{xfrac}

\allowdisplaybreaks[1]
\newcommand{\conteqn}{\nonumber\\*}
\newcommand{\neweqn}{\\}

\newcommand{\ee}[0]{\mathrm{e}}
\newcommand{\ii}[0]{i}

\newcommand{\identity}[0]{\mathbb{I}}

\newcommand{\vect}[1]{\bm{#1}}
\newcommand{\unitvect}[1]{\smash{\widehat{\vect{#1}}}\;\!\vphantom{#1}}

\newcommand{\projpm}{\projector{\pm\vect{p}}}

\newcommand{\transform}[1]{\xrightarrow{#1}}

\newcommand{\adjoint}[1]{\smash{\overline{#1}}\vphantom{#1}}

\DeclareMathOperator{\Tr}{Tr}

\newcommand{\definedby}{\equiv}

\newcommand{\deltaLDmLDn}[0]{{\delta^{{\mu}{\nu}}}}
\newcommand{\energySi}[1]{{E^{{\alpha}}(#1)}}
\newcommand{\energySj}[1]{{E^{{\beta}}(#1)}}

\newcommand{\ffElectricSi}[1][Q^2]{{G_{{E}}(#1)}}
\newcommand{\ffMagneticConv}[1][Q^2]{{G^{{\text{Conv.}}}_{{M}}(#1)}}
\newcommand{\ffMagneticPEVA}[1][Q^2]{{G^{{\text{PEVA}}}_{{M}}(#1)}}
\newcommand{\ffMagnetic}[1][Q^2]{{G_{{M}}(#1)}}
\newcommand{\ffMagneticSi}[1][Q^2]{{G_{{M}}(#1)}}

\newcommand{\ffDiracSi}[1][Q^2]{{F_{{1}}(#1)}}
\newcommand{\ffPauliSi}[1][Q^2]{{F_{{2}}(#1)}}
\newcommand{\ffStarDirac}[1][Q^2]{{F^{{*}}_{{1}}(#1)}}
\newcommand{\ffStarPauli}[1][Q^2]{{F^{{*}}_{{2}}(#1)}}

\newcommand{\gammaLCk}[0]{{\gamma^{{k}}}}
\newcommand{\gammaLUfive}[0]{{\gamma^{{5}}}}
\newcommand{\gammaLUfour}[0]{{\gamma^{{4}}}}
\newcommand{\gammaLUm}[0]{{\gamma^{{\mu}}}}
\newcommand{\gammaLUn}[0]{{\gamma^{{\nu}}}}

\newcommand{\indSdbl}[0]{{u_{\indSpr}}}

\newcommand{\indSi}[0]{{\alpha}}

\newcommand{\indSj}[0]{{\beta}}

\newcommand{\indSnesa}[0]{{n^*_1}}
\newcommand{\indSnesb}[0]{{n^*_2}}

\newcommand{\indSnsa}[0]{{N^*_1\!}}
\newcommand{\indSnsb}[0]{{N^*_2\!}}

\newcommand{\indSprsa}[0]{{p^*_1}}
\newcommand{\indSprsb}[0]{{p^*_2}}
\newcommand{\indSpr}[0]{{p}}

\newcommand{\indSsing}[0]{{d_{\indSpr}}}

\newcommand{\indexLCk}[1]{{{#1}^{{k}}}}
\newcommand{\indexLDm}[1]{{{#1}^{{\mu}}}}
\newcommand{\indexLDn}[1]{{{#1}^{{\nu}}}}
\newcommand{\interpOi}[2]{{\chi_{{i}}(#2)}}
\newcommand{\interpPEVAOip}[2]{{\chi_{{#1}\,{i'}}(#2)}}
\newcommand{\interpPEVAOi}[2]{{\chi_{{#1}\,{i}}(#2)}}

\newcommand{\interpoptPEVASi}[2]{{\phi^{{\alpha}\,{}}_{{#1}}(#2)}}
\newcommand{\interpoptPEVASj}[2]{{\phi^{{\beta}\,{}}_{{#1}}(#2)}}
\newcommand{\interpoptbarPEVASi}[2]{{\adjoint{\phi}^{{\alpha}\,{}}_{{#1}}(#2)}}

\newcommand{\massSi}[0]{{m^{{\alpha}}}}
\newcommand{\massSj}[0]{{m^{{\beta}}}}

\newcommand{\projector}[2][]{{\Gamma^{{#1}}_{\!{#2}}}}

\newcommand{\sigmaLUmLUn}[0]{{\sigma^{{\mu}{\nu}}}}

\newcommand{\spinorSi}[1][s]{{u^{{\alpha}}{(p, #1)}}}

\newcommand{\spinorpbarSi}[1][s']{{\adjoint{u}^{{\alpha}}{(p', #1)}}}
\newcommand{\spinorpbarSj}[1][s']{{\adjoint{u}^{{\beta}}{(p', #1)}}}

\newcommand{\threecfSjSi}[6][]{{\mathcal{G}^{{3}}_{{#1}}(#2\,;#3\,,#4\,;#5\,,#6\,;\indSi{}\transform{}\indSj{})}}

\newcommand{\twocfprojPEVASi}[2]{{G(#1\,;#2\,;\indSi{})}}

\newcommand{\vectorcurrentLUm}[1][]{{j^{{\mu}}_{{#1}}}}

\title{Structure and transitions of nucleon excitations via parity-expanded variational analysis}

\ShortTitle{Structure and transitions of nucleon excitations}

\author{\speaker{Finn~M.~Stokes}$\,^{ab}$, Waseem~Kamleh$^a$ and Derek~B.~Leinweber$^a$\\
    \llap{$^a$}Special Research Centre for the Subatomic Structure of
          Matter,\\Department of Physics, University of Adelaide, South
          Australia 5005, Australia\\%
    \llap{$^b$}J\"ulich Supercomputing Centre, Institute for Advanced Simulation,\\
          Forschungszentrum J\"ulich, J\"ulich D-52425, Germany\\
    E-mail: \email{f.stokes@fz-juelich.de}}

\abstract{
    The recently-introduced Parity Expanded Variational Analysis (PEVA)
    technique allows for the isolation of baryon eigenstates on the lattice at
    finite momentum free from opposite-parity contamination. We find that this
    technique introduces a statistically significant correction in extractions
    of the electromagnetic form factors of the ground state nucleon. It also
    allows first extractions of the elastic and transition form factors of
    nucleon excitations on the lattice. We present the electromagnetic elastic
    form factors and helicity amplitudes of two odd-parity excitations of
    the nucleon. These results provide valuable insight into the structure of
    these states, and allow for a connection to be made to quark-model states
    in this energy region.}

\FullConference{37th International Symposium on Lattice Field Theory - Lattice2019\\
	    16-22 June 2019\\
		Wuhan, China}

\begin{document}
\section{Introduction\label{sec:excitations:introduction}}
In lattice QCD, instead of the unstable finite-width resonances of nature, we
observe a tower of stable excitations. These eigenstates are associated
with the physical resonances in a non-trivial manner.
Understanding the structure of the states observed in Lattice QCD will enable
predictions of the infinite-volume observables of nature via effective field
theory techniques~\cite{Liu:2015ktc,Liu:2016uzk} or an extension of the
Lellouch-L{\"u}scher formalism~\cite{Lellouch:2000pv}.

Investigating the structure of excited states in lattice QCD is recognised as
an important frontier in the field. Progress has already been made in the meson
sector~\cite{Owen:2015fra,Briceno:2015dca}. Here we tackle the more challenging
problem of calculating such quantities in the baryon sector.

By using local three-quark operators on the lattice, both the
CSSM~\cite{Mahbub:2012ri} and
the Hadron Spectrum Collaboration (HSC)~\cite{Edwards:2012fx}
observe two low-lying odd-parity states in the resonance regimes of the
\(N^*(1535)\)\index{N*1535@\(N^*(1535)\)} and
\(N^*(1650)\)\index{N*1650@\(N^*(1650)\)}.
In the following we summarise our recent results on the elastic form factors of
the ground state nucleon~\cite{Stokes:2018emx}, these two odd-parity states, and the lowest-lying
even-parity state accessible through the same
operators~\cite{Stokes:2019zdd}. In addition, we present
preliminary results on the transition form factors for the two odd-parity
states. These results were made possible through the development of the PEVA
technique~\cite{Menadue:2013kfi}.

\section{Parity Expanded Variational Analysis (PEVA)}
The process of extracting elastic from factors of baryonic excited states via
the PEVA technique is presented in full in Ref~\cite{Stokes:2018emx}. We
provide here a brief summary of this process and the generalisations required
to handle transition matrix elements.

We begin with a basis of \(n\) conventional spin-\nicefrac{1}{2} operators
\(\left\{\interpOi{}{x}\right\}\) that couple to the states of interest.
Adopting the Pauli representation, we introduce the PEVA
projector~\cite{Menadue:2013kfi}
\begin{equation}
    \projpm \definedby \frac{1}{4} \left(\identity + {\gammaLUfour}\right)
    \left(\identity \pm \ii {\gammaLUfive} {\gammaLCk} \indexLCk{\unitvect{p}}\right),
\end{equation}
and construct a set of basis operators
\begin{subequations}
\begin{align}
    \interpPEVAOi{\pm\vect{p}}{x} &\definedby \projpm \, \interpOi{}{x}\,,\neweqn
    \interpPEVAOip{\pm\vect{p}}{x} &\definedby \pm \projpm \, \gammaLUfive{} \, \interpOi{}{x}\,.
\end{align}
\end{subequations}


We then seek an optimised set of operators \(\interpoptPEVASi{\pm\vect{p}}{x}\)
that each couple strongly to a single energy eigenstate \(\indSi\). These
optimised operators are constructed as linear combinations of the basis
operators by solving a generalised
eigenvalue problem as detailed in Ref.~\cite{Menadue:2013kfi}.

We can then construct the eigenstate-projected
two-point correlation function
\begin{align}
    \twocfprojPEVASi{\vect{p}}{t} &\definedby
      \Tr\left(\sum_{\vect{x}} \ee^{-\ii \vect{p}\cdot\vect{x}}
      \braket{\Omega|\,\interpoptPEVASi{\pm\vect{p}}{x}\,
        \interpoptbarPEVASi{\pm\vect{p}}{0}\,|\Omega} \right),
\end{align}
and the three point correlation functions
\begin{align}
    \threecfSjSi[\pm]{\vectorcurrentLUm[CI]{}\!}{\vect{p}'\!}{\vect{p}}{t_2}{t_1}
    &\definedby \sum_{\vect{x}_1,\vect{x}_2} \ee^{-\ii \vect{p}'\cdot\vect{x}_2} \,
        \ee^{\ii (\vect{p}' - \vect{p})\cdot\vect{x}_1}
    \braket{\Omega|\,\interpoptPEVASj{\pm\vect{p}'}{x_2}\,
        \vectorcurrentLUm[CI]{}(x_1)\,\interpoptbarPEVASi{+\vect{p}}{0}\,|\Omega},
\end{align}
where \(\vectorcurrentLUm[CI]{}(x)\) is the \(O(a)\)-improved~\cite{Martinelli:1990ny}
conserved vector current\index{vector current} \(\vectorcurrentLUm[CI]{}(x)\)
used in Ref.~\cite{Boinepalli:2006xd},
inserted with a three-momentum transfer \(\vect{q} = \vect{p}' - \vect{p}\).

For the elastic case, this choice of current gives the matrix element
\begin{align}
    &\braket{\indSi\,; p'\,; s' | \,\vectorcurrentLUm[CI]{}(0)\, | \indSi\,; p\,; s}\conteqn
    &\qquad= \sqrt{\frac{\massSi}{\energySi{\vect{p}}}} \,
              \sqrt{\frac{\massSi}{\energySi{\vect{p}'}}} \;
              \spinorpbarSi{}
    \left(\gammaLUm{} \, \ffDiracSi
        - \frac{\sigmaLUmLUn{}\,\indexLDn{q}}{2 \massSi{}} \, \ffPauliSi\right)
    \spinorSi{}\,,\label{eqn:formfactors:elasticmatrixelement}
\end{align}
where \(Q^2 = \vect{q}^2 - {\left(\energySi{\vect{p}'} - \energySi{\vect{p}}\right)}^2\),
and \(\ffDiracSi\) and \(\ffPauliSi\) are the
Dirac and Pauli form factors.
The matrix element can be extracted by taking
appropriate ratios of the three- and two-point correlation functions. The Sachs
electromagnetic form factors
\begin{equation}
    \ffElectricSi{} \definedby \ffDiracSi{} - \frac{Q^2}{{\left(2\massSi{}\right)}^2} \, \ffPauliSi{}\text{ and }
    \ffMagneticSi{} \definedby \ffDiracSi{} + \ffPauliSi{}
\end{equation}
can then be extracted by taking linear combinations of the matrix elements.

We now move on to the transition form factors. 
The relevant matrix element is~\cite{OwenThesis}
\begin{align}
   &\braket{\indSj^{-}\,; p'\,; s' | \,\vectorcurrentLUm[CI]{}(0)\, | \indSi^{+}\,; p\,; s}
   = \sqrt{\frac{\massSi}{\energySi{\vect{p}}}} \,
              \sqrt{\frac{\massSj}{\energySj{\vect{p}'}}} \conteqn
   &\qquad \times \spinorpbarSj{}
   \left(\left(\deltaLDmLDn{} - \frac{\indexLDm{q}\indexLDn{q}}{q^2}\right) \gammaLUn{}\gammaLUfive{}\, \ffStarDirac
        - \frac{\sigmaLUmLUn{}\,\indexLDn{q}}{\massSj{} - \massSi{}} \, \gammaLUfive\, \ffStarPauli\right)
    \spinorSi{}\,,\label{eqn:formfactors:matrixelement}
\end{align}
where \(\ffStarDirac\) and \(\ffStarPauli\) are
Dirac-\index{form factors!Dirac} and Pauli-like\index{form factors!Pauli}
transition form factors. We can then take ratios and linear combinations
to obtain the transverse helicity amplitude
\begin{align}
    A_{\nicefrac{1}{2}}\!\left(Q^2\right) &\definedby
    2\sqrt{\frac{Q^2 + {\left(\massSj{} - \massSi{}\right)}^2}{8\massSi\left(\massSj{}^2 - \massSi{}^2\right)}}
    \left(\ffStarDirac{} + \ffStarPauli{}\right).
\end{align}

\section{Ground state nucleon}
We study the extraction of the elastic form factors of the ground state nucleon
in detail in Ref.~\cite{Stokes:2018emx}. This analysis is performed on the PACS-CS
\((2+1)\)-flavour full-QCD ensembles~\cite{Aoki:2008sm}, made available through the
ILDG~\cite{Beckett:2009cb}. In the paper we demonstrate the
efficacy of variational analysis techniques in general, and
PEVA specifically, at controlling excited-state contaminations in the
electric form factor. Both the PEVA and conventional variational analysis show
clear and clean plateaus, supporting previous work demonstrating the utility of
variational analysis in calculating baryon matrix
elements~\cite{Dragos:2016rtx, Owen:2012ts}.

\begin{figure}
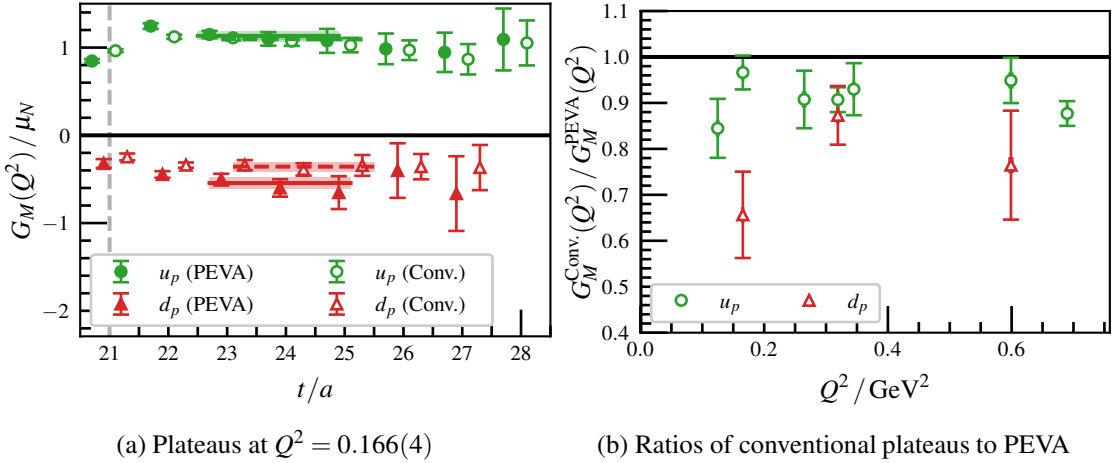

\begin{subfigure}[b]{0.48\textwidth}
    \centering
    \input{state0_fit_p000_pp100_k13781_GM.pgf}
    \caption{\label{fig:groundstate:gm:plateau}Plateaus at \(Q^2 = 0.166(4)\)}
\end{subfigure}
\begin{subfigure}[b]{0.48\textwidth}
    \centering
    \input{state0_qs_ratio_k13781_GM.pgf}
    \caption{\label{fig:groundstate:gm:ratio}Ratios of conventional plateaus to PEVA}
\end{subfigure}
    \caption{Comparison of conventional and PEVA extractions of \(\ffMagnetic{}\)
    for the ground-state nucleon
    at \(m_\pi = \SI{156}{\mega\electronvolt}\). Results are contributions for
    single quarks of unit charge from the doubly represented quark sector (\(u_p\))
    and the singly represented quark sector (\(d_p\)).}
\end{figure}

Here we focus on the particular case of the magnetic form factor, where we
found evidence that the conventional analysis is contaminated by opposite-parity
states. In Fig.~\ref{fig:groundstate:gm:plateau} we plot a comparison of magnetic
form factor plateaus produced by a conventional variational analysis (using an initial
basis of \(n=8\) operators), with an equivalent extraction via the PEVA technique
(with the basis parity-expanded to \(2n=16\) operators). We see a significant difference in the
plateaus extracted by the two techniques for the singly represented quark sector.
If we take the correlated ratio of the extracted values, as shown for a range
of kinematics in Fig.~\ref{fig:groundstate:gm:ratio}, we see a consistent
underestimation of the value by the conventional analysis. This shows
\(\sim 20\%\) underestimation of the magnitude of the contributions
to the magnetic form factor from the singly represented quark flavour in the
conventional analysis.

The difference between the two analyses is that the PEVA
approach provides additional interpolator degrees of freedom to improve the
ground state interpolating field at finite momentum. As such it is clear that the difference
between the two extractions is from contaminating states that are present in
the conventional analysis but removed by the parity expansion. As such, the PEVA
technique is critical for precision measurements of nucleon form factors.

\section{Excitations}
With our local three-quark operators, we observe two low-lying odd-parity
eigenstates in the resonance regimes of the \(N^*(1535)\) and \(N^*(1650)\).
In Ref.~\cite{Stokes:2019zdd}, we investigate the elastic form factors of these
states. There we find that opposite parity contaminations have a large effect
on both the electric and magnetic form factors, and the PEVA technique is
critical for even a qualitatively correct extraction.

We focus our investigation at heavier pion masses, where these lattice states
lie below the relevant two-particle scattering thresholds on the finite volume.
At these masses, we find that these states look remarkably similar to constituent
quark model predictions for the \(N^*(1535)\) and \(N^*(1650)\).

\begin{figure}
    \centering
    \input{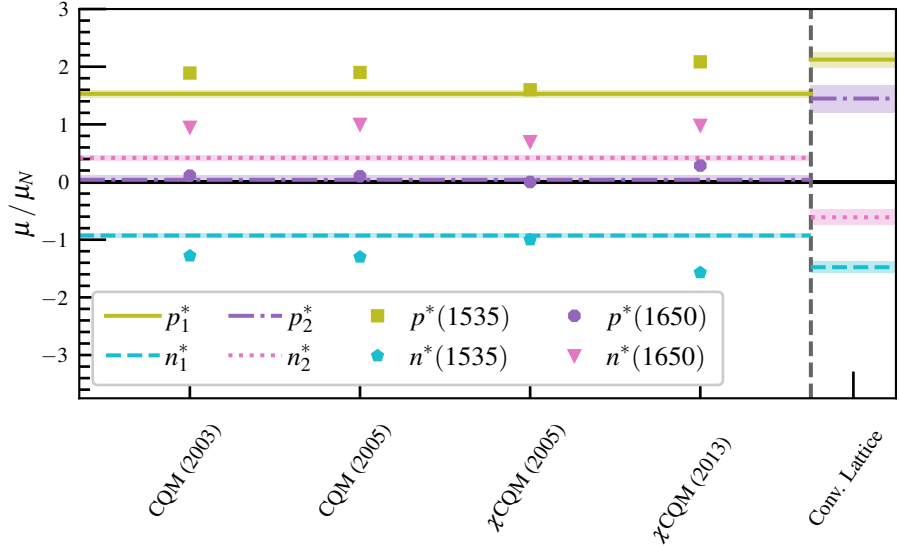}
    \caption{\label{fig:negpar:mm}Comparison between lattice
    calculations of the magnetic moments of two odd-parity nucleon
    excitations at \(m_{\pi} = \SI{702}{\mega\electronvolt}\)
    and quark model predictions~\cite{Chiang:2002ah,Liu:2005wg,Sharma:2013rka} for the \(N^*(1535)\)%
    \index{N*1535@\(N^*(1535)\)} and \(N^*(1650)\) resonances%
    \index{N*1650@\(N^*(1650)\)}. The shaded bands on the left-hand side of the
    plot indicate the
    magnetic moments calculated via the PEVA technique in lattice QCD,
    and symbols denote the quark model predictions.
    Lattice calculations of the magnetic moments using
    conventional parity projection are plotted to the right of the vertical
    dashed line.}
\end{figure}

We find the size of these lattice eigenstates to be similar to the ground state
nucleons. As shown in Fig.~\ref{fig:negpar:mm}, their magnetic moments
agree well with constituent quark model predictions for the continuum states.

\begin{figure}[t]
    \centering
    \input{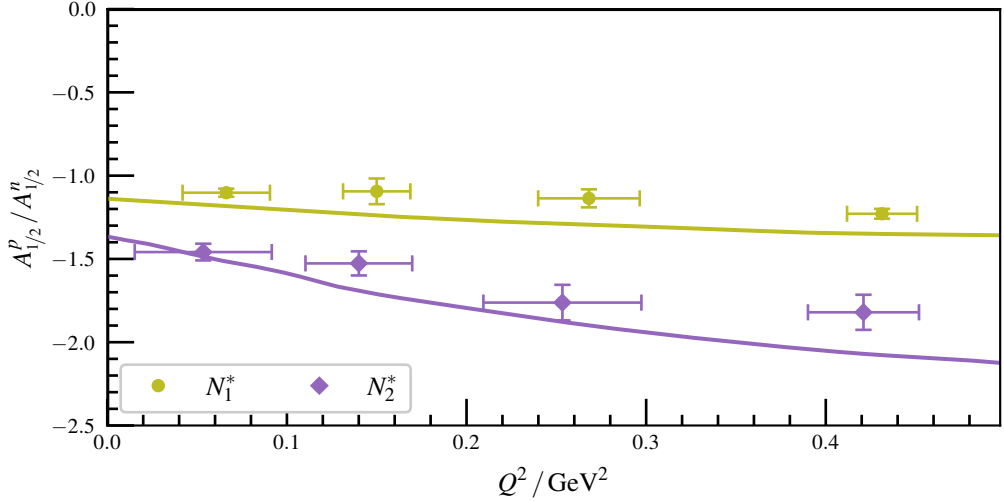}
    \caption{\label{fig:negpar:A}Constituent quark model
    predictions~\cite{Capstick:1994ne} (lines) and PEVA extractions (points) of
    the ratio of proton to neutron helicity amplitudes 
    at \(m_{\pi} = \SI{702}{\mega\electronvolt}\).}
\end{figure}

We also present here preliminary results for the transition form factors from
the ground state to each of these two lattice eigenstates. These results will
be presented in more detail in an upcoming paper. In Fig.~\ref{fig:negpar:A}
we compare the ratio of the transverse helicity amplitudes of the proton and
neutron to constituent quark model predictions. Taking this ratio allows us to
cancel out some of the model dependence of the constituent quark model result.
Similar to the magnetic moments above, we compare the first negative-parity
lattice excitation (\(\indSnsa\)) to a quark model prediction for the
\(N^*(1535)\), and the second lattice excitation (\(\indSnsb\))
to a prediction for the \(N^*(1650)\).
We once again find good agreement between the structure of the lattice eigenstates
at the heavier pion masses and the constituent quark model prediction.

We see strong agreement between the lattice eigenstates and constituent quark model
predictions at these heavier pion masses. This suggests that while the dynamics are
much more complicated at the physical point, and the constituent quark model
alone does not appear to give a good description of these resonances in nature,
as the pion mass increases the constituent quark model describes the
excitations rather well.
The agreement of the lighter state with the constituent-quark-model \(N^*(1535)\)
is consistent with predictions from Hamiltonian Effective Field Theory
(HEFT)~\cite{Liu:2015ktc}. However, the agreement of the heavier state with the
constituent-quark-model \(N^*(1650)\) suggests that future HEFT studies should
explore the incorporation of two bare basis states associated with the two
different localised states observed herein.

In Ref.~\cite{Stokes:2019zdd}, we also investigate the lowest-lying even-parity
excitation of the nucleon observed on the lattice. We find that it has a charge radius approximately
\(30\%\) larger than the ground state, and a remarkably similar
magnetic moment to the ground state. This is consistent with the state being a radial excitation of
the ground-state nucleon as seen in Refs.~\cite{Roberts:2013ipa}.

\section{Conclusion}
The PEVA technique is critical to correctly extracting the form factors of
nucleon excitations on the lattice. Such extractions give us insight
into the structure of these finite-volume states. In addition, even for the
ground state we found evidence that the
conventional analysis was contaminated by opposite-parity states. For
the kinematics considered here, we observe \(\sim 20\%\)
underestimation of the contributions to the magnetic form
factor from the singly represented quark flavour at lighter pion masses.
All these results make it clear that the PEVA
technique is critical for precision measurements of nucleon form factors
and for any study of the structure of nucleon excitations.

\section*{Acknowledgements}
This research was undertaken with the assistance of resources from the
Phoenix HPC service at the University of Adelaide, the National
Computational Infrastructure (NCI), which is supported by the Australian
Government, and by resources provided by the Pawsey Supercomputing Centre
with funding from the Australian Government and the Government of Western
Australia. These resources were provided through the National Computational
Merit Allocation Scheme and the University of Adelaide partner share. This
research is supported by the Australian Research Council (ARC) through grants
no.\ DP140103067, DP150103164, LE160100051, and DP190102215.

\bibliographystyle{JHEPMod}
\bibliography{referencemod}

\end{document}